\documentclass[journal]{IEEEtran}

\usepackage{cite}

\ifCLASSINFOpdf
\else
   \usepackage[dvips]{graphicx}
\fi
\usepackage[cmex10]{amsmath}
\usepackage{amssymb}
\usepackage{balance}

\newtheorem{theorem}{Theorem}
\newtheorem{corollary}{Corollary}

\newcommand{\ato}{\overset{\mathrm{a.s.}}{\to}}
\newcommand{\aeq}{\overset{\mathrm{a.s.}}{=}}

\begin{document}
%
\title{Convolutional Approximate Message-Passing}
%
%
%

\author{Keigo~Takeuchi,~\IEEEmembership{Member,~IEEE}
\thanks{
The author was in part supported by the Grant-in-Aid 
for Scientific Research~(B) (JSPS KAKENHI Grant Number 18H01441), Japan. 
}
\thanks{K.~Takeuchi is with the Department of Electrical and Electronic Information Engineering, Toyohashi University of Technology, Aichi 441-8580, Japan (e-mail: takeuchi@ee.tut.ac.jp).}
}

\markboth{IEEE transactions on signal processing letters,~Vol.~, No.~,}%
{Takeuchi: Convolutional Approximate Message-Passing}
%

\IEEEpubid{0000--0000/00\$00.00~\copyright~2020 IEEE}


\maketitle

\begin{abstract}
This letter proposes a novel message-passing algorithm for signal recovery 
in compressed sensing. The proposed algorithm solves the disadvantages of 
approximate message-passing (AMP) and orthogonal/vector AMP, and realizes 
their advantages. AMP converges only in a limited class of sensing 
matrices while it has low complexity. Orthogonal/vector AMP requires a 
high-complexity matrix inversion while it is applicable for a wide class of 
sensing matrices. The key feature of the proposed algorithm is the so-called 
Onsager correction via a convolution of messages in all preceding iterations 
while the conventional message-passing algorithms have correction terms that 
depend only on messages in the latest iteration. Thus, the proposed algorithm 
is called convolutional AMP (CAMP). Ill-conditioned sensing matrices are 
simulated as an example in which the convergence of AMP 
is not guaranteed. Numerical simulations show that CAMP can improve the 
convergence property of AMP and achieve high performance comparable to 
orthogonal/vector AMP in spite of low complexity comparable to AMP.
\end{abstract}


%
\IEEEpeerreviewmaketitle

\section{Introduction}
\IEEEPARstart{A}{pproximate} message-passing (AMP)~\cite{Donoho09} is a 
low-complexity algorithm of signal recovery in compressed 
sensing~\cite{Donoho06,Candes06}. When 
the sensing matrix has independent and identically distributed (i.i.d.) 
zero-mean sub-Gaussian elements~\cite{Bayati11,Bayati15}, AMP was proved to be 
asymptotically Bayes-optimal in a certain region of the compression rate.    
However, AMP fails to converge when the sensing matrix is 
ill-conditioned~\cite{Rangan191} or has non-zero mean~\cite{Caltagirone14}. 

To solve this convergence issue of AMP, orthogonal AMP (OAMP)~\cite{Ma17} 
and vector AMP (VAMP)~\cite{Rangan192} were proposed. OAMP and VAMP are 
equivalent to each other. The Bayes-optimal version of OAMP/VAMP was 
originally proposed by Opper and Winther~\cite{Opper05}. 
OAMP/VAMP was proved to be asymptotically Bayes-optimal when the sensing 
matrix is orthogonally invariant~\cite{Rangan192,Takeuchi192}. 
However, OAMP/VAMP has high complexity unless the singular-value 
decomposition (SVD) of the sensing matrix can be computed efficiently. 

This letter proposes a novel message-passing (MP) algorithm that solves both 
the convergence issue of AMP and the complexity issue of OAMP/VAMP. The 
proposed MP uses the same matched filter as AMP while OAMP/VAMP utilizes a 
linear minimum mean-square error (LMMSE) filter. Furthermore, it performs the 
so-called Onsager correction via a convolution of messages in all preceding 
iterations while AMP and OAMP/VAMP have correction terms that depend only 
on messages in the latest iteration. Thus, the proposed MP is called 
convolutional AMP (CAMP). 

The tap coefficients in the convolution are determined so as to realize the 
asymptotic Gaussianity of the estimation errors of CAMP. For that purpose, 
they are defined such that a general error model proposed in 
\cite{Takeuchi191} contains the error model of CAMP asymptotically. 
Since the asymptotic Gaussianity in the general error model has been proved 
for any orthogonally invariant sensing matrix~\cite{Takeuchi191}, 
the estimation errors of CAMP are asymptotically Gaussian-distributed. 
Numerical simulations for ill-conditioned sensing matrices show that CAMP 
can achieve performance comparable to 
OAMP/VAMP in spite of complexity comparable to AMP. 

\IEEEpubidadjcol

\section{Measurement Model}
Consider the $M$-dimensional linear measurements 
$\boldsymbol{y}\in\mathbb{R}^{M}$ of an unknown $N$-dimensional sparse 
signal vector $\boldsymbol{x}\in\mathbb{R}^{N}$,  
\begin{equation} \label{model}
\boldsymbol{y} 
= \boldsymbol{A}\boldsymbol{x} + \boldsymbol{w}, 
\quad 
\boldsymbol{w}\sim\mathcal{N}(\boldsymbol{0},\sigma^{2}\boldsymbol{I}_{M}). 
\end{equation}
In (\ref{model}), $\boldsymbol{A}\in\mathbb{R}^{M\times N}$ denotes a known 
sensing matrix. The vector $\boldsymbol{w}$ is additive white Gaussian 
noise (AWGN) with covariance $\sigma^{2}\boldsymbol{I}_{M}$. The triple 
$(\boldsymbol{A}, \boldsymbol{x}, \boldsymbol{w})$ is independent random 
variables. For simplicity, the signal vector $\boldsymbol{x}$ is assumed to 
have i.i.d.\ elements with 
zero mean and unit variance. Furthermore, the power normalization 
$N^{-1}\mathbb{E}[\|\boldsymbol{A}\|^{2}]=1$ is assumed.  

An important assumption is the right-orthogonal invariance of 
$\boldsymbol{A}$: In the SVD 
$\boldsymbol{A}=\boldsymbol{U}\boldsymbol{\Sigma}\boldsymbol{V}^{\mathrm{T}}$, 
the $N\times N$ orthogonal matrix $\boldsymbol{V}$ is independent of 
$\boldsymbol{U}\boldsymbol{\Sigma}$ and Haar-distributed~\cite{Hiai00}. 
This class of matrices contains zero-mean i.i.d. Gaussian matrices. 

As an additional technical assumption, the empirical eigenvalue distribution 
of $\boldsymbol{A}^{\mathrm{T}}\boldsymbol{A}$ converges almost surely to 
a deterministic distribution with a compact support in the large system 
limit, in which $M$ and $N$ tend to infinity while the compression rate 
$\delta=M/N$ is kept ${\cal O}(1)$. Let $\mu_{k}$ denote the $k$th moment of 
the empirical eigenvalue distribution, 
\begin{equation} \label{moment} 
\mu_{k} = \frac{1}{N}\mathrm{Tr}\left(
 \boldsymbol{\Lambda}^{k}
\right), 
\end{equation}
with $\boldsymbol{\Lambda}=\boldsymbol{\Sigma}^{\mathrm{T}}\boldsymbol{\Sigma}$. 
The technical assumption implies that any moment $\mu_{k}$ converges 
almost surely in the large system limit. In particular, 
the power normalization $N^{-1}\mathbb{E}[\|\boldsymbol{A}\|^{2}]=1$ implies 
$\mu_{1}\ato1$ in the large system limit. 

\section{Convolutional AMP}
\subsection{Algorithm}
The so-called Onsager correction is used to guarantee the asymptotic 
Gaussianity of the estimation errors before thresholding in each iteration of 
MP. 
The Onsager correction in AMP depends only on a message in the latest 
iteration. While AMP is a low-complexity algorithm, the Onsager correction 
in AMP fails to guarantee the asymptotic Gaussianity, with the only exception 
of zero-mean i.i.d.\ sensing matrices~\cite{Bayati11,Bayati15}. 

The proposed CAMP has Onsager correction applicable to all 
right-orthogonally invariant sensing matrices. The correction term is 
a convolution of messages in all preceding iterations. Thus, the proposed 
MP is called convolutional AMP.   

Let $\boldsymbol{x}_{t}\in\mathbb{R}^{N}$ denote an estimator of 
$\boldsymbol{x}$ in iteration~$t$ of CAMP. The estimator $\boldsymbol{x}_{t}$ 
is recursively given by 
\begin{equation}
\boldsymbol{x}_{t+1} = f_{t}(\boldsymbol{x}_{t}+\boldsymbol{A}^{\mathrm{T}}
\boldsymbol{z}_{t}), 
\end{equation}
\begin{equation} \label{z}
\boldsymbol{z}_{t} 
= \boldsymbol{y} - \boldsymbol{A}\boldsymbol{x}_{t} 
+ \sum_{\tau=0}^{t-1}\xi_{\tau}^{(t-1)}g_{t-\tau-1}^{(1)}\boldsymbol{z}_{\tau}, 
\end{equation}
\begin{equation} \label{xi}
\xi_{t}^{(t')} 
= \prod_{\tau=t}^{t'}\left\langle
 f'_{\tau}(\boldsymbol{x}_{\tau}
 +\boldsymbol{A}^{\mathrm{T}}\boldsymbol{z}_{\tau})
\right\rangle, 
\end{equation}
with $\boldsymbol{x}_{0}=\boldsymbol{0}$. 
In the CAMP, $\{f_{t}:\mathbb{R}\to\mathbb{R}\}$ are a sequence of  
Lipschitz-continuous thresholding functions. 
For any function $f:\mathbb{R}\to\mathbb{R}$, 
$f(\boldsymbol{v})$ represents the element-wise application of 
$f$ to a vector $\boldsymbol{v}$, i.e.\ $[f(\boldsymbol{v})]_{n}
=f([\boldsymbol{v}]_{n})$. The notional convention $\sum_{\tau=0}^{-1}\cdots
=\boldsymbol{0}$ is used in (\ref{z}). 
The notation $\langle \boldsymbol{v} 
\rangle=N^{-1}\sum_{n=1}^{N}v_{n}$ denotes the arithmetic mean of the elements of 
$\boldsymbol{v}=(v_{1},\ldots,v_{N})^{\mathrm{T}}$. 
The CAMP reduces to conventional AMP in the case of 
$g_{0}^{(1)}=\delta^{-1}$ and $g_{t}^{(1)}=0$ for all $t>0$.   

To define the tap coefficients $\{g_{t}^{(1)}\}$ in the CAMP, consider a 
discrete-time dynamical system $\{g_{t}^{(k)}: k, t=0,1,\ldots\}$, 
\begin{equation} \label{g0}
g_{0}^{(k)} = \mu_{k+1} - \mu_{k}, 
\end{equation}
\begin{equation} \label{g1}
g_{1}^{(k)} 
= g_{0}^{(k)} - g_{0}^{(k+1)} + g_{0}^{(1)}\mu_{k+1}, 
\end{equation}
\begin{IEEEeqnarray}{rl}
g_{t}^{(k)} 
=& g_{t-1}^{(k)} - g_{t-1}^{(k+1)}  
+ \sum_{\tau=1}^{t-1}g_{t-\tau-1}^{(1)}\left(
 g_{\tau}^{(k)} - g_{\tau-1}^{(k)}
\right) \nonumber \\
&+ g_{t-1}^{(1)}\mu_{k+1}
\quad \hbox{for $t\geq2$,}  \label{gt}  
\end{IEEEeqnarray}
where $\mu_{k}$ denotes the $k$th moment~(\ref{moment}) of the empirical 
eigenvalue distribution of $\boldsymbol{A}^{\mathrm{T}}\boldsymbol{A}$. 

In a practical implementation, the moment sequence should be replaced by 
the asymptotic one in the large system limit. This replacement implies that 
the complexity to compute $\{g_{t}^{(1)}\}$ can be independent of the system 
size if the asymptotic eigenvalue distribution of 
$\boldsymbol{A}^{\mathrm{T}}\boldsymbol{A}$ has a closed-form expression. 

\begin{table}
\begin{center}
\caption{
Complexity in $M\leq N$ and the number of iterations~$t$. 
}
\label{table1}
\begin{tabular}{|c|c|c|}
\hline 
AMP & OAMP/VAMP & CAMP \\ 
\hline 
${\cal O}(tMN)$ & ${\cal O}(M^{2}N+tMN)$ 
& ${\cal O}(tMN + t^{2}M)$ \\
\hline 
\end{tabular}
\end{center}
\end{table}

The computational complexity of the CAMP, AMP, and OAMP/VAMP is compared 
in Table~\ref{table1}. The complexity of AMP is dominated by 
matrix-vector multiplication. The first term for OAMP/VAMP is the worst-case 
complexity of the SVD of $\boldsymbol{A}$. The second term for the CAMP is 
due to computation of the Onsager correction term. As long as the number of 
iterations~$t$ is much smaller than $M$ and $N$, the complexity of the CAMP 
is comparable to that of AMP.

\subsection{State Evolution}
The tap coefficients in the CAMP have been determined so as to guarantee the 
asymptotic Gaussianity of the estimation errors. 
The author~\cite{Takeuchi191} proposed a general error model and used 
state evolution (SE) to prove that the estimation error before thresholding is 
asymptotically Gaussian-distributed in the general error model. 
To prove the asymptotic Gaussianity of the estimation error 
$\boldsymbol{h}_{t}=\boldsymbol{x}_{t}+\boldsymbol{A}^{\mathrm{T}}
\boldsymbol{z}_{t} - \boldsymbol{x}$ before the thresholding $f_{t}$, thus, 
it is sufficient to show that the error model of the CAMP is included into 
the general error model. 

Let $\boldsymbol{q}_{t+1}=f_{t}(\boldsymbol{x}+\boldsymbol{h}_{t})
-\boldsymbol{x}$ denote the estimation error after the thresholding. 
According to the definition of the general error model~\cite{Takeuchi191}, 
define $\boldsymbol{b}_{t}=\boldsymbol{V}^{\mathrm{T}}\tilde{\boldsymbol{q}}_{t}$, 
$\boldsymbol{m}_{t}=\boldsymbol{V}^{\mathrm{T}}\boldsymbol{h}_{t}$, and 
\begin{equation} \label{q_tilde}
\tilde{\boldsymbol{q}}_{0} = \boldsymbol{q}_{0}, \quad 
\tilde{\boldsymbol{q}}_{t} = \boldsymbol{q}_{t} - \xi_{t-1}\boldsymbol{h}_{t-1}  
\end{equation}
for $t>0$, 
where $\xi_{t}$ is an abbreviation of $\xi_{t}^{(t)}$ given in (\ref{xi}). 
Then, $\boldsymbol{m}_{t}$ satisfies the following equation: 
\begin{IEEEeqnarray}{rl} 
\boldsymbol{m}_{t} 
=& (\boldsymbol{I}_{N}-\boldsymbol{\Lambda})
(\boldsymbol{b}_{t} + \xi_{t-1}\boldsymbol{m}_{t-1})
+ \boldsymbol{\Sigma}^{\mathrm{T}}\boldsymbol{U}^{\mathrm{T}}\boldsymbol{w} 
\nonumber \\ 
&+ \sum_{\tau=0}^{t-1}\xi_{\tau}^{(t-1)}g_{t-\tau-1}^{(1)}\left(
 \boldsymbol{m}_{\tau} - \boldsymbol{b}_{\tau} 
 - \xi_{\tau-1}\boldsymbol{m}_{\tau-1}
\right), \label{m}
\end{IEEEeqnarray}
with $\boldsymbol{m}_{t}=\boldsymbol{0}$ for all $t<0$.

\begin{IEEEproof}[Proof of (\ref{m})]
From the definitions of $\boldsymbol{m}_{t}$ and 
$\boldsymbol{h}_{t}$, we use the SVD $\boldsymbol{A}=\boldsymbol{U}
\boldsymbol{\Sigma}\boldsymbol{V}^{\mathrm{T}}$ to have 
\begin{equation} \label{m_tmp}
\boldsymbol{m}_{t} = \boldsymbol{V}^{\mathrm{T}}\boldsymbol{q}_{t} 
+ \boldsymbol{\Sigma}^{\mathrm{T}}\boldsymbol{U}^{\mathrm{T}}\boldsymbol{z}_{t}. 
\end{equation}
Left-multiplying (\ref{z}) by $\boldsymbol{\Sigma}^{\mathrm{T}}
\boldsymbol{U}^{\mathrm{T}}$ and substituting (\ref{model}) and (\ref{m_tmp}), 
we obtain 
\begin{IEEEeqnarray}{rl}
\boldsymbol{m}_{t} 
=& (\boldsymbol{I}_{N}-\boldsymbol{\Lambda})\boldsymbol{V}^{\mathrm{T}}
\boldsymbol{q}_{t} 
+ \boldsymbol{\Sigma}^{\mathrm{T}}\boldsymbol{U}^{\mathrm{T}}\boldsymbol{w} 
\nonumber \\ 
&+ \sum_{\tau=0}^{t-1}\xi_{\tau}^{(t-1)}g_{t-\tau-1}^{(1)}\left(
 \boldsymbol{m}_{\tau} - \boldsymbol{V}^{\mathrm{T}}\boldsymbol{q}_{\tau} 
\right). 
\end{IEEEeqnarray}
Using (\ref{q_tilde}) and the definitions of $\boldsymbol{b}_{t}$ and 
$\boldsymbol{m}_{t}$, we arrive at (\ref{m}). 
\end{IEEEproof}

For $\tau=0,1,\ldots$ and $\tau'=0,\ldots,\tau$, define 
\begin{equation}
g_{\tau', \tau}^{(k)} 
= \frac{1}{N}\sum_{n=1}^{N}
\frac{\partial[\boldsymbol{\Lambda}^{k}\boldsymbol{m}_{\tau}]_{n}}
{\partial[\boldsymbol{b}_{\tau'}]_{n}}. 
\end{equation}
When $g_{\tau',\tau}^{(0)}=0$ holds for all $\tau'$ and $\tau$, 
the general error model in \cite{Takeuchi191} includes the error model of the 
CAMP. The following theorem implies that the 
inclusion is correct in the large system limit. Thus, the asymptotic 
Gaussianity of the estimation errors is guaranteed in the CAMP. 
\begin{theorem} \label{theorem1}
For all $\tau=0,1,\ldots$ and $\tau'=0,\ldots,\tau$, the almost sure 
convergence $g_{\tau', \tau}^{(0)}\ato0$ holds in the large system limit.  
\end{theorem}
\begin{IEEEproof}
The proof is by induction to show
\begin{enumerate}
\item $g_{\tau',\tau}^{(0)}\ato0$, \label{statement1}
\item the almost sure convergence of $\xi_{\tau}$ to a constant, 
\label{statement2}
\item Let $\tilde{g}_{0}^{(k)}=g_{\tau,\tau}^{(k)}$ and 
$\tilde{g}_{\tau',\tau}^{(k)}=g_{\tau',\tau}^{(k)}/\xi_{\tau'}^{(\tau-1)}$ for 
$\tau'<\tau$. $\tilde{g}_{\tau',\tau}^{(k)}$ depends on 
$\tau$ and $\tau'$ only through $\tau-\tau'$.  
\label{statement3}
\end{enumerate} 

According to \cite[Theorem~1]{Takeuchi191}, the statement~\ref{statement2})  
follows from the statement~\ref{statement1}). Thus, we only focus on 
the first and last statements.  
For $\tau=0$, we use (\ref{m}) to obtain $g_{0,0}^{(0)} = \mu_{0} - \mu_{1}\ato0$, 
because of $\mu_{0}=1$ and $\mu_{1}\ato1$. 

For some $t$, assume the three statements for all $\tau<t$ and $\tau'\leq\tau$. 
We shall prove the first and last statements for $\tau=t$. 

We first prove the statement~\ref{statement3}). 
For $t'=t$ and $t'=t-1$, we use (\ref{m}) to obtain 
\begin{equation} \label{gtt}
g_{t,t}^{(k)} = \mu_{k} - \mu_{k+1}, 
\end{equation}
\begin{equation}
g_{t-1,t}^{(k)} 
= \xi_{t-1}(g_{t-1,t-1}^{(k)} - g_{t-1,t-1}^{(k+1)}) 
+ \xi_{t-1}g_{0}^{(1)}(g_{t-1,t-1}^{(k)} - \mu_{k}), 
\end{equation}
where we have used the second induction hypothesis. 
Similarly, for $t'\leq t-2$ we have 
\begin{IEEEeqnarray}{rl}
g_{t',t}^{(k)} 
=& \xi_{t-1}(g_{t',t-1}^{(k)} - g_{t',t-1}^{(k+1)}) 
+ \sum_{\tau=t'}^{t-1}\xi_{\tau}^{(t-1)}g_{t-\tau-1}^{(1)}g_{t',\tau}^{(k)}   
\nonumber \\
-& \xi_{t'}^{(t-1)}g_{t-t'-1}^{(1)}\mu_{k}   
- \sum_{\tau=t'+1}^{t-1}\xi_{\tau-1}^{(t-1)}g_{t-\tau-1}^{(1)}g_{t',\tau-1}^{(k)}.    
\end{IEEEeqnarray}

From the last induction hypothesis, 
we can define $\tilde{g}_{\tau-\tau'}^{(k)}=g_{\tau',\tau}^{(k)}/\xi_{\tau'}^{\tau-1}$ 
for $\tau<t$. Using (\ref{gtt}) and this change of variables yields  
\begin{equation} \label{g1_tilde}
\tilde{g}_{t-1,t}^{(k)} 
= \tilde{g}_{0}^{(k)} - \tilde{g}_{0}^{(k+1)} 
- g_{0}^{(1)}\mu_{k+1},  
\end{equation}
\begin{IEEEeqnarray}{rl}
\tilde{g}_{t',t}^{(k)} 
=& \tilde{g}_{t-t'-1}^{(k)} - \tilde{g}_{t-t'-1}^{(k+1)} 
+ \sum_{\tau=1}^{t-t'-1}g_{t-t'-\tau-1}^{(1)}\left(
 \tilde{g}_{\tau}^{(k)} - \tilde{g}_{\tau-1}^{(k)}
\right)  
\nonumber \\
&- g_{t-t'-1}^{(1)}\mu_{k+1}
    \label{gt_tilde}
\end{IEEEeqnarray}
for $t'\leq t-2$, with $\tilde{g}_{0}^{(k)}=g_{t,t}^{(k)}$. 
Since the right-hand sides (RHSs) depend on $t'$ and $t$ only through 
$t-t'$, we find that the statement~\ref{statement3}) holds for $\tau=t$, and 
can re-write the left-hand sides of (\ref{g1_tilde}) and (\ref{gt_tilde}) 
as $\tilde{g}_{1}^{(k)}$ and $\tilde{g}_{t-t'}^{(k)}$, respectively.  

Finally, we prove the statement~\ref{statement1}). It is sufficient to prove 
$\tilde{g}_{t'}^{(0)}\ato0$ and $\tilde{g}_{t'}^{(k)}+g_{t'}^{(k)}\ato0$ 
for all $t'=0,\ldots,t$ and $k$. The proof is by induction. For $t'=0$, 
we have $\tilde{g}_{0}^{(0)}=\mu_{0}-\mu_{1}\ato0$. 
Comparing (\ref{g0}) and $\tilde{g}_{0}^{(k)}=\mu_{k}-\mu_{k+1}$ yields 
$\tilde{g}_{0}^{(k)}=-g_{0}^{(k)}$.  

For $t'=1$, we use (\ref{g1_tilde}), $\tilde{g}_{0}^{(0)}\ato0$, and 
$\tilde{g}_{0}^{(1)}=-g_{0}^{(1)}$ to obtain 
$\tilde{g}_{1}^{(0)}\aeq - \tilde{g}_{0}^{(1)} - g_{0}^{(1)} + o(1)\ato0$. 
Furthermore, we use (\ref{g1}), (\ref{g1_tilde}), and 
$\tilde{g}_{0}^{(k)}=-g_{0}^{(k)}$ 
to find $\tilde{g}_{1}^{(k)}+g_{1}^{(k)}\ato0$.  

Assume $\tilde{g}_{t'}^{(0)}\ato0$ and $\tilde{g}_{t'}^{(k)}+g_{t'}^{(k)}\ato0$ 
for all $t'<\tau\in\{2,\ldots, t\}$, and prove $\tilde{g}_{\tau}^{(0)}\ato0$ and 
$\tilde{g}_{\tau}^{(k)}+g_{\tau}^{(k)}\ato0$. For the former statement, 
we use (\ref{gt_tilde}) and the induction hypotheses 
$\tilde{g}_{t'}^{(0)}\ato0$ and $\tilde{g}_{\tau-1}^{(1)}+g_{\tau-1}^{(1)}\ato0$ 
to obtain 
\begin{equation}
\tilde{g}_{\tau}^{(0)} 
\aeq - \tilde{g}_{\tau-1}^{(1)} 
- g_{\tau-1}^{(1)} + o(1) \ato 0. 
\end{equation}
For the latter statement, we use (\ref{gt}), (\ref{gt_tilde}), and 
the induction hypothesis $\tilde{g}_{t'}^{(k)} + g_{t'}^{(k)}\ato0$ to find  
$\tilde{g}_{\tau}^{(k)} + g_{\tau}^{(k)}\ato0$. Thus, 
$\tilde{g}_{t'}^{(0)}\ato0$ and $\tilde{g}_{t'}^{(k)}+g_{t'}^{(k)}\ato0$ hold  
for all $t'=0,\ldots,t$ and $k$. In other words, we have 
proved the statement~\ref{statement1}). 
\end{IEEEproof}

\subsection{Closed-Form Solution} 
The sequence $\{g_{t}^{(k)}\}$ may be computed by solving the discrete-time 
dynamical systems~(\ref{g0})--(\ref{gt}) numerically when the moment 
sequence $\{\mu_{k}\}$ is given. However, it is possible to obtain a 
closed-form solution of the tap coefficients $\{g_{t}^{(1)}\}$ via 
the $\eta$-transform $\eta(z)$ of the asymptotic eigenvalue distribution of 
$\boldsymbol{A}^{\mathrm{T}}\boldsymbol{A}$~\cite{Tulino04}, 
given by 
\begin{equation} \label{eta_series}
\eta(z) = \lim_{M=\delta N\to\infty}\sum_{k=0}^{\infty}\mu_{k}(-z)^{k}.  
\end{equation}

\begin{theorem} \label{theorem2} 
Let $G_{k}(y)$ denote the generating function of $\{g_{t}^{(k)}\}$ with 
respect to $t=0,1\,\ldots$, defined as 
\begin{equation} \label{Gk_y}
G_{k}(y) = \sum_{t=0}^{\infty}y^{t}g_{t}^{(k)}. 
\end{equation}
Then, $G_{1}(y)$ is implicitly given by $\eta(x_{\mathrm{s}}) = 1 - y$ 
in the large system limit, with  
\begin{equation} \label{xs}
x_{\mathrm{s}}
= \frac{y}{(1-y)\{1-yG_{1}(y)\}}. 
\end{equation}
\end{theorem}
\begin{IEEEproof}
Define the generating function of $\{g_{t}^{(k)}\}$ as 
\begin{equation} 
G(x,y) = \sum_{k=0}^{\infty}x^{k}G_{k}(y). 
\end{equation}
Theorem~\ref{theorem2} follows from the following closed-form expression of 
$G(x,y)$: 
\begin{equation} \label{G}
G(x,y) 
\aeq \frac{x\eta(-x)\{yG_{1}(y)-1\} + \eta(-x)-1}
{(1-y)\{1-yG_{1}(y)\}x + y} + o(1). 
\end{equation}

By definition, $G(x,y)$ is a polynomial of $x$ and $y$. Thus, the numerator 
of (\ref{G}) must be zero when the denominator is zero. The point 
$-x_{\mathrm{s}}$ given in (\ref{xs}) is a zero of the denominator for any $y$. 
Thus, we let the numerator at $x=-x_{\mathrm{s}}$ be zero to obtain  
$\eta(x_{\mathrm{s}})=1-y$. Thus, we arrive at Theorem~\ref{theorem2}. 

To complete the proof of Theorem~\ref{theorem2}, we shall prove (\ref{G}). 
We first derive a closed-form expression of (\ref{Gk_y}), given by 
\begin{equation} \label{Gk_y_tmp}
G_{k}(y) 
= g_{0}^{(k)} + g_{1}^{(k)}y + \sum_{t=2}^{\infty}y^{t}g_{t}^{(k)}. 
\end{equation}
Substituting (\ref{gt}) into the last term on the RHS of (\ref{Gk_y_tmp}) 
yields 
\begin{IEEEeqnarray}{l}
\sum_{t=2}^{\infty}y^{t}g_{t}^{(k)}
= \sum_{t=2}^{\infty}y^{t}g_{t-1}^{(k)} - \sum_{t=2}^{\infty}y^{t}g_{t-1}^{(k+1)}
+ \sum_{t=2}^{\infty}y^{t}g_{t-1}^{(1)}\mu_{k+1}
\nonumber \\
+ \sum_{t=2}^{\infty}y^{t}\sum_{\tau=1}^{t-1}g_{t-\tau-1}^{(1)}g_{\tau}^{(k)}
- \sum_{t=2}^{\infty}y^{t}\sum_{\tau=1}^{t-1}g_{t-\tau-1}^{(1)}g_{\tau-1}^{(k)}.  
\end{IEEEeqnarray}
For the first three terms, we have 
\begin{equation}
\sum_{t=2}^{\infty}y^{t}g_{t-1}^{(k')}
= y\sum_{t=1}^{\infty}y^{t}g_{t}^{(k')}
= yG_{k'}(y) - g_{0}^{(k')}y 
\end{equation}
for $k'=1, k, k+1$. Since the Z-transform of convolution is the product of 
Z-transforms, the last term reduces to 
\begin{IEEEeqnarray}{rl}
\sum_{t=2}^{\infty}y^{t}\sum_{\tau=1}^{t-1}g_{t-\tau-1}^{(1)}g_{\tau-1}^{(k)}
=& y^{2}\sum_{t=0}^{\infty}y^{t}\sum_{\tau=0}^{t}g_{t-\tau}^{(1)}g_{\tau}^{(k)} 
\nonumber \\ 
=& y^{2}G_{1}(y)G_{k}(y).  
\end{IEEEeqnarray} 
Similarly, for the fourth term we have 
\begin{IEEEeqnarray}{rl}
\sum_{t=2}^{\infty}y^{t}\sum_{\tau=1}^{t-1}g_{t-\tau-1}^{(1)}g_{\tau}^{(k)}
=& \sum_{t=1}^{\infty}y^{t+1}\sum_{\tau=1}^{t}g_{t-\tau}^{(1)}g_{\tau}^{(k)} 
\nonumber \\ 
=& yG_{1}(y)\left\{
 G_{k}(y) - g_{0}^{(k)}
\right\}. 
\end{IEEEeqnarray}
Using these results, as well as (\ref{g0}) and (\ref{g1}), 
we obtain the closed-form expression  
\begin{equation} \label{Gk} 
G_{k}(y) 
= \frac{\mu_{k}yG_{1}(y) - yG_{k+1}(y) + g_{0}^{(k)}}
{(1-y)\{1-yG_{1}(y)\}}. 
\end{equation}

The closed-form expression~(\ref{G}) follows from (\ref{Gk}). 
Using the $\eta$-transform~(\ref{eta_series}) yields 
\begin{equation}
G(x,y) 
\aeq \frac{\eta(-x)yG_{1}(y) - x^{-1}yG(x,y) + G(x,0)}
{(1-y)\{1-yG_{1}(y)\}} + o(1),  
\end{equation}
where we have used $G_{0}(y)\ato0$ obtained from Theorem~\ref{theorem1}. 
Applying $G(x,0)=x^{-1}\{\eta(-x)-1\}-\eta(-x)$ obtained 
from (\ref{g0}) and solving $G(x,y)$, we arrive at (\ref{G}). 
\end{IEEEproof}

The following corollary implies that the CAMP reduces to conventional AMP 
when the sensing matrix has i.i.d.\ Gaussian elements with mean proportional 
to $M^{-1/2}$. 
Thus, the CAMP has no ability to handle this non-zero mean case. 
\begin{corollary}
If $\boldsymbol{A}$ has independent Gaussian elements with mean 
$\sqrt{\gamma/M}$ and variance $(1-\gamma)/M$ for any $\gamma\in[0, 1)$, 
the CAMP is equivalent to conventional AMP. 
\end{corollary}
\begin{IEEEproof}
The R-transform $R(z)$~\cite[Section 2.4.2]{Tulino04} of the asymptotic 
eigenvalue distribution of $\boldsymbol{A}^{\mathrm{T}}\boldsymbol{A}$ is 
given by 
\begin{equation}
R(z) = \frac{\delta}{\delta-z}. 
\end{equation}
Using Theorem~\ref{theorem2} and the following relationship between 
the $\eta$ and R transforms: 
\begin{equation}
\eta(z) = \frac{1}{1+zR(-z\eta(z))}, 
\end{equation}
we obtain
\begin{equation}
1-y = \frac{1}{1+\delta x_{\mathrm{s}}\{\delta+x_{\mathrm{s}}(1-y)\}^{-1}}, 
\end{equation}
where $x_{\mathrm{s}}$ is given by (\ref{xs}). Substituting (\ref{xs}) and 
solving $G_{1}(y)$, we arrive at $G_{1}(y)=\delta^{-1}$. 

From the definition~(\ref{Gk_y}), we find $g_{0}^{(1)}=\delta^{-1}$ and 
$g_{t}^{(1)}=0$ for all $t>0$. This implies that the update rule~(\ref{z}) 
reduces to that corresponding to conventional AMP. 
\end{IEEEproof} 

The following corollary is utilized in numerical simulations. 
\begin{corollary} \label{corollary2}
If $\boldsymbol{A}$ is orthogonally invariant and has non-zero singular values 
$\sigma_{0}\geq\cdots\geq\sigma_{M-1}>0$ satisfying condition number 
$\kappa=\sigma_{0}/\sigma_{M-1}\geq1$, $\sigma_{m}/\sigma_{m-1}=\kappa^{-1/(M-1)}$, 
and $\sigma_{0}^{2}=N(1-\kappa^{-2/(M-1)})/(1-\kappa^{-2M/(M-1)})$, then 
$g_{t}^{(1)}=g_{t}+C/(\kappa^{2}-1)$ holds for all $t$, with 
\begin{equation}
g_{t} = \sum_{\tau=0}^{t-1}h_{t-\tau}g_{\tau} - h_{t+1}, \quad 
g_{0} = -h_{1}, 
\end{equation}  
\begin{equation}
h_{t}= \frac{C^{t-1}}{t!} - \frac{C^{t}}{(t+1)!}, 
\quad C=\frac{2}{\delta}\ln\kappa. 
\end{equation}
\end{corollary}
\begin{IEEEproof}
Since $\mu_{k}=N^{-1}\sigma_{0}^{2k}(1-\kappa^{-2kM/(M-1)})/(1-\kappa^{-2k/(M-1)})$ 
holds for all $k>0$, we use (\ref{eta_series}) and 
$N(1-\kappa^{-a/(M-1)})\to\delta^{-1}a\ln\kappa$ for any $a\in\mathbb{R}$ to find 
\begin{IEEEeqnarray}{rl}
\eta(z) =& 1 + \sum_{k=1}^{\infty}(-z)^{k}\left\{
 \frac{C}{(1-\kappa^{-2})}
\right\}^{k}\frac{(1-\kappa^{-2k})}{kC}
\nonumber \\
=& 1 
- \frac{1}{C}\ln\left\{
 \frac{\delta(\kappa^{2}-1)+2\kappa^{2}z\ln\kappa}
 {\delta(\kappa^{2}-1)+2z\ln\kappa}
\right\}, 
\end{IEEEeqnarray}
where the second equality follows from   
$\ln(1+x)=\sum_{k=1}^{\infty}(-1)^{k-1}k^{-1}x^{k}$ for all $|x|<1$. 
Using Theorem~\ref{theorem2} yields 
\begin{equation}
G_{1}(y) 
= \frac{C}{(\kappa^{2}-1)(1-y)} 
+ \frac{1}{y} + \frac{C}{(1-y)(1-e^{Cy})}. 
\end{equation}
It is an exercise to confirm that the generating function of $g_{t}$ in 
Corollary~\ref{corollary2} is equal to the sum of the second and last terms. 
Thus, Corollary~\ref{corollary2} holds. 
\end{IEEEproof}

\section{Numerical Simulation} 
The CAMP is compared to AMP and OAMP/VAMP in terms of the mean-square error 
(MSE) in signal recovery. As an example of ill-conditioned sensing matrices 
in Corollary~\ref{corollary2}, $\boldsymbol{A}=\mathrm{diag}\{\sigma_{0},
\ldots,\sigma_{M-1}\}\boldsymbol{H}$ is considered for $M\leq N$, with $\sigma_{m}$ denoting 
the $m$th singular value in Corollary~\ref{corollary2}. The $M$ rows of 
$\boldsymbol{H}\in\mathbb{R}^{M\times N}$ are selected uniformly and randomly 
from the rows of the $N\times N$ Hadamard orthogonal matrix. 
  
We assume the Bernoulli-Gaussian (BG) prior: Each signal element takes $0$ with 
probability $1-\rho$. Otherwise, it is sampled from the zero-mean 
Gaussian distribution with variance $\rho^{-1}$. We use the soft 
thresholding~\cite{Donoho09}  
\begin{equation}
f_{t}(x)
= \left\{
 \begin{array}{cl}
 x-\theta_{t} & \hbox{for $x\geq \theta_{t}$,} \\
 0 & \hbox{for $x\in(-\theta_{t}, \theta_{t})$,} \\
 x + \theta_{t} & \hbox{for $x\leq-\theta_{t}$.}
 \end{array}
\right.  
\end{equation}
For the sensing matrix in Corollary~\ref{corollary2}, we have no SE results of 
the CAMP or AMP for designing the threshold $\theta_{t}$. Thus, the 
threshold $\theta_{t}$ is fixed to a constant $\theta$ over all iterations, 
which was optimized via an exhaustive search. 

\begin{figure}[t]
\begin{center}
\includegraphics[width=\hsize]{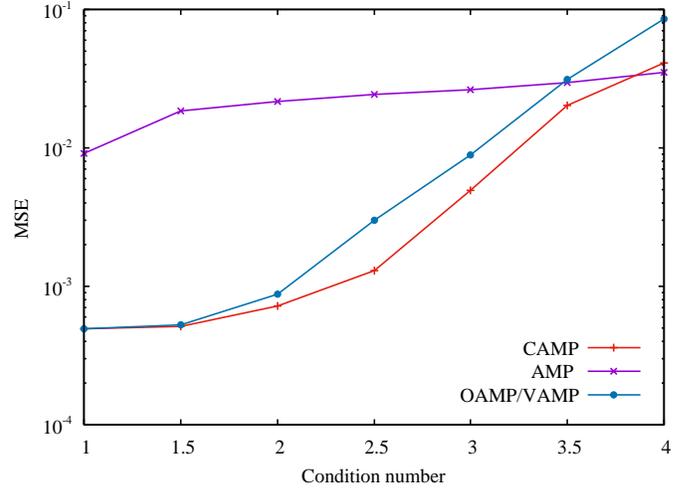}
\end{center}
\caption{
MSE versus the condition number~$\sigma_{0}/\sigma_{M-1}$ for signal density 
$\rho=0.1$, $100$ iterations, $M=614$, $N=2^{10}$, and $1/\sigma^{2}=30$~dB. 
}
\label{fig1} 
\end{figure}


Figure~\ref{fig1} shows the MSEs of the CAMP, AMP, and OAMP/VAMP estimated from 
$10^5$ independent trials. 
The CAMP outperforms AMP and achieves the MSEs comparable to OAMP/VAMP. 
The inferior performance of AMP is due to a bad convergence property   
of AMP. Using a large threshold $\theta$ improves the convergence property. 
Exhaustive search of $\theta$ implied that larger thresholds are required for 
AMP to converge than for the other algorithms. Thus, we conclude that CAMP 
improves the convergence property of AMP. 

\ifCLASSOPTIONcaptionsoff
  \newpage
\fi



\bibliographystyle{IEEEtran}
\bibliography{IEEEabrv,kt-spl2019}
\end{document}